\nonstopmode
\documentclass{entcs} 
\usepackage{my}
\usepackage{amssymb,amsmath}
\usepackage{wasysym}
\usepackage{prooftree}
\usepackage{subfigure,url}
\usepackage{qsymbols}
\usepackage{mathpazo}

\usepackage{pstricks}
\usepackage{pst-node}
\usepackage{graphicx}

\newcommand{\Act}{\textit{Act}}

\newcommand{\Val}{ \mathcal{V }}

\newcommand{\nil}{\mathbf{0}}

\def\prove#1#2{\setbox1=\hbox{$#1$}\setbox2=\hbox{$#2$}
\ifdim\wd1<\wd2
      \dimen1=\wd2
\else\dimen1=\wd1
\fi
\baselineskip=2pt
\vtop{\hbox{\vbox{\hbox to \dimen1{\hfill\vbox{\box1\kern4pt}\hfill}
      \hbox{\vbox{\hrule height 0.4pt width \dimen1\kern2pt}}}}
      \hbox to \dimen1{\hfill\vbox{\kern2pt\box2}\hfill}}}

 \def \RuleusingThree[#1]#2#3{\prooftree #2\justifies#3 \using{(#1)}\endprooftree}
 \def \Ruleusing[#1]#2{\@ifnextchar\bgroup {\RuleusingThree[#1]{#2}}
	{\prooftree \justifies #2 \using{(#1)} \endprooftree}}
 \def \RulenotusingTwo#1#2{\prooftree #1 \justifies #2 \endprooftree}
 \def \Rulenotusing#1{\@ifnextchar\bgroup {\RulenotusingTwo{#1}}
	{\prooftree \justifies #1 \endprooftree}}
 \def \prove{\proofrulebaseline=1.8ex
	\abovedisplayskip12pt\belowdisplayskip12pt
	\abovedisplayshortskip8pt\belowdisplayshortskip8pt
	\@ifnextchar[{\Ruleusing}{\Rulenotusing}}

\newcommand{\meet}{\sqcap}
\newcommand{\join}{\sqcup}

\newcommand{\Proc}{{\cal P}\kern-1.5pt\textit{r}}

\newcommand{\procgenhigh}{{\cal P}\kern-1.5pt\textit{r}_{H}}
\newcommand{\procgeninhigh}{{\cal P}\kern-1.5pt\textit{r}_{I_{H}}}

\newcommand{\Names}{{\cal N}}  
\newcommand{\Conames}{\overline{{\cal N}}}
\newcommand{\highnames}{\mathcal{H}}

\def \Def {\setbox138=\hbox{$\,=\,$}\setbox139=\hbox{\scriptsize{$\Delta$}}%
\copy138\kern-.5\wd138\kern-.5\wd139\raise6pt\hbox{\copy139}\kern-.5\wd139\kern.5\wd138}
\newcommand{\union}{\cup }

\def\InPoint #1(#2).{#1(#2)\hspace*{.5pt}.\hspace*{1.5pt}}
\def\OutPoint #1<#2>.{\overline{#1}\Use{#2}\hspace*{.5pt}.\hspace*{1.5pt}}
\makeatletter
\def \In #1(#2){\@ifnextchar{.}{\InPoint {#1}({#2})}{#1(#2)}}
\def\Out#1<#2>{\@ifnextchar{.}{\OutPoint {#1}<{#2}>}{\overline{#1}\Use{#2}}}
\makeatother
\newcommand{\Nil}{\mathbf{0}}
\newcommand{\New}[1]{\nu#1}
\newcommand{\Par}{\hspace*{1pt}{\mid}\hspace*{1pt}}

\def\Pair<#1,#2>{\langle#1,#2\rangle}
\newcommand{\Use}[1]{\langle #1\rangle }
\newcommand{\fn}[1]{\textit{fn}( #1) }
\newcommand{\Choice}[2]{\hbox{$\sum$}_{#1 \in #2}\hspace*{1pt}}

\newcommand{\Agent}[2]{#1 [ \Vec{#2} ]}

\newcommand{\uptohigh}[1]{\dRel{\sim}{#1}}
\newcommand{\highprocesses}{\Pi}

\newcommand{\coname}[1]{\overline{#1}}
\newcommand{\subst}[2]{\{#2/#1\}}
\newcommand{\Set}[1]{\{\hspace*{1pt}#1\hspace*{1pt}\}}

\def \longharpoon#1{\psset{unit=1pt,linewidth=0.35pt}%
\psline{cc-cc}(0,1.5)(.9#1,1.5)%
\hspace*{.9#1}\pscurve{cc-cc}(0,1.5)(-1.1,2)(-2.25,3.7)\hspace*{.1#1}}

\def \Vec#1{\setbox155=\hbox{$#1$}%
\leavevmode\copy155\kern-.95\wd155
\raise\ht155\hbox{\longharpoon{\wd155}}}

\newcommand{\Comment}[1]{}

\newcommand{\lts}[1]{\,\stackrel{#1}{\longrightarrow}\,}
\newcommand{\ltss}[1]{\,{\stackrel{#1}{\longrightarrow}}\kern-1.5pt\raise5pt\hbox{\scriptsize $*$}\,}
\newcommand{\wlts}[1]{\,\stackrel{#1}{\Rightarrow}\,}

\renewcommand{\lts}[1]{\MyArrow{#1}}

\newcommand{\bndc}{{\sf BNDC}}
\newcommand{\pbndc}{\textsf{P{-}BNDC}}
\newcommand{\wbndc}{\textsf{W{-}BNDC}}
\def\Bisi{ \mathcal{S}}
\def \bisimilarity{\approx}
\def \bisimilarityuptohigh{\approx_\highnames}
\def \bisimilaritytohigh{\approx_{h}}
\def \bisimilaritytohigh2{\approx_{H_{I}}}
\def \notbisi{\not\approx}

\newcommand{\notout}{\mathsf{F}}
\newcommand{\yesout}{\mathsf{T}}
\newcommand{\typetop}{\top}
\newcommand{\typebot}{\bot}

\def \Der #1 |- #2 : (#3,#4,#5){#1 \vdash #2 : (#4,#5)}
\def \DerFT #1 |- #2 : #3 {#1 \vdash #2 : #3}

\newcommand{\gentype}{\mathsf{B}}

\makeatletter
 \def \RuleusingThree[#1]#2#3{\prooftree #2\justifies#3 \using{(#1)}\endprooftree}
 \def \Ruleusing[#1]#2{\@ifnextchar\bgroup {\RuleusingThree[#1]{#2}}
	{\prooftree \justifies #2 \using{(#1)} \endprooftree}}
 \def \RulenotusingTwo#1#2{\prooftree #1 \justifies #2 \endprooftree}
 \def \Rulenotusing#1{\@ifnextchar\bgroup {\RulenotusingTwo{#1}}
	{\prooftree \justifies #1 \endprooftree}}
 \def \prove{\proofrulebaseline=1.625ex
	\abovedisplayskip12pt\belowdisplayskip12pt
	\abovedisplayshortskip8pt\belowdisplayshortskip8pt
	\@ifnextchar[{\Ruleusing}{\Rulenotusing}}
\makeatother

\usepackage{pstricks}

\def \dRel#1#2{\setbox155=\hbox{\scriptsize $#1$}\setbox156=\hbox{\scriptsize $#2$}%
\,\raise-1pt\copy156%
\kern-.5\wd156\kern-.5\wd155%
\raise3pt\copy155%
\kern-.5\wd155\kern.5\wd156\,}

\def \MyArrow#1{%
\setbox155=\hbox{\scriptsize $#1$}%
\setbox156=\hbox{$\longrightarrow$}%
\,\raise-1pt\copy156%
\kern-.5\wd156\kern-.5\wd155%
\raise4pt\copy155%
\kern-.5\wd155\kern.5\wd156\,}

\newcommand{\subj}[1]{\mathit{subj}(#1)}

\newcommand{\lattice}[2]{(\mathcal{#1},#2)}
\newcommand{\mylattice}{\mathcal{L}}

\newcommand{\ptnil}{\textsc{\small Nil}}

\newcommand{\ptsum}{\textsc{\small Sum}}
\newcommand{\ptcom}{\textsc{\small Comp}}
\newcommand{\ptrec}{\textsc{\small Rec}}
\newcommand{\ptsub}{\textsc{\small Sub}}
\newcommand{\ptres}{\textsc{\small Restr}}
\newcommand{\ptnorm}{\textsc{\small Norm}}

\newcommand{\ltsinput}{\textsc{\small Input}}
\newcommand{\ltsoutput}{\textsc{\small Output}}
\newcommand{\ltsparright}{\textsc{\small Par Right }}
\newcommand{\ltsparleft}{\textsc{\small Par Left}}
\newcommand{\ltsparcommone}{\textsc{\small Par Comm$_1$}}
\newcommand{\ltsparcommtwo}{\textsc{\small Par Comm$_2$}}
\newcommand{\ltsres}{\textsc{\small Restr}}
\newcommand{\ltsrec}{\textsc{\small Rec}}

\newcommand{\ruletypenil}{\textsc{\small Nil}}
\newcommand{\ruletypepar}{\textsc{\small Par}}
\newcommand{\ruletypesum}{\textsc{\small Sum}}
\newcommand{\ruletypesub}{\textsc{\small SubType}}
\newcommand{\ruletypeinput}{\textsc{\small Input}}
\newcommand{\ruletypeinputone}{\textsc{\small Input-1}}
\newcommand{\ruletypeinputtwo}{\textsc{\small Input-2}}
\newcommand{\ruletypeoutput}{\textsc{\small Output}}
\newcommand{\ruletypeoutputone}{\textsc{\small Output-1}}
\newcommand{\ruletypeoutputtwo}{\textsc{\small Output-2}}
\newcommand{\ruletyperestr}{\textsc{\small Restr}}
\newcommand{\ruletyperec}{\textsc{\small Rec}}

\newcommand{\PI}{$\pi$-calculus}

\begin{document}

\def\lastname{van Bakel \& Vigliotti}
\begin{frontmatter}
 \title{Note on a simple type system for non-interference}
 \author{Steffen van Bakel \and Maria Grazia Vigliotti}
 \address{Department of Computing, Imperial College London, \\180 Queen's Gate, London SW7 2BZ, UK}
 \thanks[myemail]{Email: \href{mailto:svb@doc.ic.ac.uk }{ \texttt{\normalshape svb@doc.ic.ac.uk}}, \href{mailto:mgv98@doc.ic.ac.uk }{ \texttt{\normalshape
 mgv98@doc.ic.ac.uk}}}

\begin{abstract} We consider CCS with value
passing and elaborate a notion of noninterference for the process calculi,
which matches closely that of the programming language. 
The idea is to view channels as information carriers rather than as
``events'', so that emitting a secret on output channel can be considered safe, while inputting a secret may lead to some kind of leakage. 
This is in contrast with the standard notion of noninterference for the process calculi where any causal dependency of low-level action from any high-level action is forbidden.
\end{abstract}


\end{frontmatter}

\section*{Introduction}
\label{sec:intro}

In recent years {\em secure information flow} has attracted a great deal of interest, spurred on by the spreading of mobile devices and nomadic computation, and has been studied in some depth for both programming languages and process calculi. 
In this paper we shall speak of the ``language-based approach'' when referring to programming languages and of the ``process-algebraic approach'' when referring to process calculi.

The language-based approach is concerned with the avoidance of secret information leakage or corruption through the execution of programs, i.e.~with the security properties of {\em confidentiality} and {\em integrity}. 
The property of confidentiality, which appears to be the most studied, is usually formalised via the notion of {\em non-interference}, meaning that secret inputs should not have an effect on public outputs, since this could allow -in principle- a public user to reconstruct sensitive information.
Non-interference may be achieved in various ways: via program analysis, type systems, using semantics equivalencies, the implementation of security policies, etc. 
In most cases the languages are equipped with a type system or some other tool to enforce the compliance of programs to the desired security property.

In the process-algebraic approach the focus is on the notion of {\em external observer}, who ideally has nothing to do with the specification and implementation of a given system, and should not be able to infer any secret by interacting with it.
The process-algebraic approach is concerned with secret events not being revealed while processes communicate, i.e.~actions that involve sensitive or confidential data should have no effect on public actions.
 
Also, in process algebra, many non-interference properties are formalised in a way similar to programming languages, i.e.~using program analysis, type systems, using semantics equivalencies. 
In the last few years a variety of properties have been proposed for process calculi, mostly based on trace equivalence or bisimulation, ranging from the simple property of {\em Non-deducibility on Composition} to more complicated ones (see~\cite{Focardi-Gorrieri'01} for a review).

Methods for static detection of insecure processes have not been largely studied for process calculi. 
In~\cite{Hennessy-Riely'02,Hennessy'04} type systems which characterise a non-inference property have been proposed for the $\pi$-calculus. 
More sophisticated type systems have been extensively studied in~\cite{Honda-Vasc-el'00,Honda-Yoshida'02} for variants of the $\pi$-calculus, which combine the control of security with other correctness concerns. 
More recently, Crafa and Rossi proposed in~\cite{Crafa-Rossi'05} a simple security type system for the $\pi$-calculus, which consists essentially of a simplification of that used by Hennessy \cite{Hennessy'04}, ensuring the absence of explicit information flows. 
All those type systems include specific analysis on the values passed on a channel. 

Pottier~\cite{Pottier'02} proposed a very simple view on non-interference via a type system for the $\pi$-calculus which does not involve any extra typing information on the values passed over channels. 
The great appeal of this type system is its simplicity in characterising non-interference only; in fact, Pottier calls this system 'simple', and we will use his terminology in this paper. 
The limitation of Pottier's work, with respect to the 'simple type system' is the lack of a robust semantic notion of non-interference. 
In this paper we will address this issue specifically.

In process algebraic approach, differently from language based security, no distinction is made between input events and output events, neither at the level of semantics definitions of security not at the level of type systems.
In this paper we aim to address two issues: 

\begin{enumerate}

\item study the relationship between those type systems and the {\em semantics-based approach} in process calculi~\cite{Focardi-Gorrieri'01,Focardi-el'05,Focardi'02};

\item to define a notion of non-interference which matches closely the one in the language-based approach. 
In other words, the basic idea is to view channels as information carriers, so that emitting a secret on an output channel can be considered safe, while 
inputting a secret may lead to some kind of leakage. 

\end{enumerate}

As for the first issue, the notion of {\em Persistent Non-deducibility on Composition} developed for CCS~\cite{Focardi-el'05,Focardi'02} has shown to be quite natural, also because it preserves the notion of non-interference of the language-based approach in the process-algebraic approach~\cite{Focardi-el'05}. 
In this paper we will show that the `simple type system' can be adapted to standard CCS and that it characterises the semantic notion of Persistent Non-deducibility on Composition. 
This means that any typeable process is persistently deducible on composition. 
We will show that there exist processes that are considered secure according the notion of persistence, yet that are not typeable. 
Therefore, the set of typeable processes according to Pottier's type system is strictly smaller than the class of processes included in Persistent Non-deducibility on Composition relation.

We consider CCS here instead of the $\pi$-calculus because we wish to focus on the specific issues of non-interference in the simplest model possible. 
It is clear that our work could be easily extended to the $\pi$-calculus, with little extra effort.
As for the second issue, we modify the `simple type system' so that the notion of non-interference matches closely that of programming languages. That is to view channels as information carriers rather than as ``events'', so that the process $\In a_h(x).\Out b_l<e>$, which emits on a low channel a value received on a high channel, is considered insecure, while $\Out a_h<v>.\Out b_l<v>$, which emits successively a value $v$ on a high channel and on a low channel, is considered secure. 
The second example would not be be typeable in the `simple type system' nor would it be considered secure with the standard semantic notions of non-interference. \\

The rest of the paper is organised as follows: in section \ref{sec:niccs} we introduce CCS; in section \ref{equiv-security} we introduce the notion of equivalence-based security; in section \ref{simple-type-system} we adapt the simple type system to CCS and we show that the every typeable process is secure according to the 
Persistent Non-deducibility on Composition. 
Finally, in section \ref{refined-type-system} we introduce our refined type system and elaborate on a semantics notion of non-interference based on the idea that only high-level inputs are critical for the definition of non-interference. 
Conclusions follow.

\section{ CCS}
\label{sec:niccs}

We will consider a variant of CCS with value passing, with two main differences from standard presentation:
\begin{enumerate}
\item We assume the existence of a lattice $\lattice{L}{\leq}$, which expresses the security level of channels. Greek letters $`s,`t,`r \ldots$ and $\ell$ range over $\mylattice$.
The language CCS we consider is typed in the sense that we explicitly incorporate the security level of the channel in the syntax of the language. 
 \item We consider the value passing CCS -though value passing could be encoded with infinite choice operator \cite{Milner'89}- without if-then-else operator as in ~\cite{Focardi-el'05}. We prefer to consider CCS with value passing in order to emphasise the different role of input and output; yet the if-then-else operator can be encoded in CCS 
 \cite{Milner'89} and therefore is not essential in the current presentation. 
\end{enumerate}

\begin{definition}
Let $\Names$ be a enumerable set of names and $\Conames$ an enumerable set of conames.
 We use the usual conventions for input $\In a(x)$ and output $\Out a<e>$.
 The enumerable set of {\em variables} is ranged over by $x,y,z \ldots$, and the set of {\em values} $\Val$ is ranged over by $e$; we will assume that $ (\Names \cup \Conames) \cap \Val =\emptyset $. 

The syntax of (typed) {\em process prefixes}, ranged over by $`a,`b,`g$, is given by:
 \[ \begin{array}{rcl}
`a &::=& \In {a_\ell}(x) \mid \Out a_\ell<e> 
\end{array} \]
where $\ell$ is taken from a lattice $\lattice{L}{\leq}$of security levels. 

The set $\Proc$ of processes, ranged over by $P,Q$, is given by the grammar:
\[\begin{array}{rcl}
P, Q & ::= & \Nil \mid \Choice{i}{I}`a_{i}.P_i \mid P {\Par} Q \mid (\New a_\ell)\,P \mid \Agent{A}{e}.\\
\end{array} \]
where $I$ is a finite index set.
\end{definition}


The informal meaning of process is standard: {\em choice operator} $ \Choice{i}{I}`a_{i}. 
P_i$ represents the non-deterministic choice among different processes; {\em parallel composition} $P \Par Q$ represent processes running together, possibly in an interleaving fashion; {\em restriction} $ (\New a_\ell)\,P $ makes the name $a_\ell$ local to the process $P$.

\begin{definition}[Notions and Conventions]
\begin{itemize}

\item The notion of free and bound names in $P$
is standard, taking into account that $(\New a)\,P$ is the only binding operator. With $n(P)$ we mean the set of names in $P$.

\item For an {\em Agent} $\Agent{A}{a}$ we assume the existence of identifier $A$ such that a process $P$ can be associated to that identifier, written $\Agent{A}{x} \Def P$ when $\fn{P} \subseteq \Set{x_1, \ldots x_n}$.


\item We assume that prefixes with the same channel name have the same security level i.e.~if $\In a_{\ell}(x).P $ and $\Out a_{\ell'}<e>.Q $ then $\ell=\ell'$. 

\item We write $P\subst{x}{e}$ ($P\subst{x}{a}$) for the standard replacement of every occurrence of $x$ in $P$ by the value $e$ (the name $a$). 

\item An element of the set of actions {\Act} is defined as $\Act \Def \Set{ae \mid a \in \Names \cup \Conames, \, e\in \Val } \cup \Set {`t}$; the Greek letters $`a, `b \ldots $ will range over \Act.

\item We define $\subj{\In a_{\ell}(x)}=a_{\ell}=\subj{\Out a_{\ell}<e>}$ and $\subj{`t}= `t$. 
\end{itemize}

\end{definition}

\begin{definition}[Operational Semantics]
The relation $\lts{} \subseteq \Proc \times \Act\times \Proc $, written $P \lts{`a} P'$, is defined by
: 
\[ 
\begin{array}{rl}
(\ltsinput): & 
	\prove[`a_j = \In a_{\ell}(x)]
		{\Choice{i}{I}`a_i.P_i \lts{ae}P_j\subst{x}{e}}
\\[3mm]
(\ltsoutput): &
	\prove[`a_j = \Out a_{\ell}<e>]
		{\Choice{i}{I}`a_i.P_i \lts{\coname{a}e}P_j}
\\[3mm]
(\ltsres): & 
	\prove[b\neq \subj{`a}]
		{ P \lts{`a} P'}
		{(\New{b})P \lts{`a} (\New{b}) P'}
\\[6mm]
(\ltsrec): &
 	\prove[P\Def A(\Vec{x})]
		{P\Vec{\subst{x}{b}} \lts{`a} P'}
		{ \Agent{A}{b} \lts{`a} P'} 

\end{array}
 \begin{array}{rl}
(\ltsparleft): & 
	\prove{P \lts{`a} P'}
		{P \Par Q \lts{`a} P' \Par Q}
\\[6mm] 
(\ltsparright): & 
	\prove{P \lts{`a} P'}
		{Q \Par P \lts{`a}Q \Par P'}
\\[6mm]
(\ltsparcommone): & 
	\prove{P \lts{ae}P' \quad Q \lts{\coname{a}e}Q' }
		{P \Par Q \lts{`t} P'\Par Q'} 

\\[6mm]
(\ltsparcommtwo): & 
	\prove{P \lts{\coname{a}e}P' \quad Q \lts{ae}Q' }
		{P \Par Q \lts{`t} P'\Par Q'} 
\end{array}\]
\end{definition}
We adopt the usual notational conventions.
We write $\ltss{`t}$ for the reflexive and transitive closure of $\lts{`t}$. 
We define define $P \wlts{`a} P'$ as $P \ltss{`t}\lts{`a}\ltss{`t} P'$ and $P \wlts{\hat{`a}} P'$ as $P \wlts{`a} P'$ if $`a \not= `t$ or $P \ltss{`t}P'$ otherwise. Thus $P \wlts{`t}{P'}$ requires at least one $`t$-transition while
$P\wlts{\hat{`t}}{P'}$ allows for the empty move.

\section{Equivalence-based security}\label{equiv-security}
In this section, we shall examine previous definitions of equivalence-based security that aim to capture the notions of non-interference. There are many different definitions, based on semantics equivalencies \cite{Focardi-Gorrieri'01}. We consider in this paper only non-interference bisimilarity for two reasons: (1) these equivalencies are very common in the literature \cite{Crafa-Rossi'05,Pottier'02,Focardi'02,Focardi-el'05,Boudol-Castellani'02}, etc, and (2) there are well-established proof-methods to show when processes are equivalent. 
We shall first consider Bisimulation-based Non-Deducibility on Compositions (\bndc) followed by Persistent Bisimulation-based Non-Deducibility on Compositions (\pbndc).

\begin{definition}[Weak Bisimulation]
A symmetric binary relation $\Bisi \subseteq \Proc\times \Proc $ is a {\em weak bisimulation} if $P \Bisi Q$ implies, for all $a \in \Act$: 
 \begin{itemize}
 \item whenever $P \lts{`a}P'$ then there exists a $Q'$ such that $Q\wlts{\hat{`a}}Q'$ and $P' \Bisi Q'$.
 \end{itemize}
Two processes $P$ and $Q$ are {\em weakly bisimilar}, written $P
\bisimilarity Q$, if for some weak bisimulation $\Bisi$, $P \Bisi Q$.
\end{definition}

It is well known that $\bisimilarity$ is both the largest bisimulation and an equivalence relation.
 
In this section, we will assume --without loss of generality-- that the lattice of security levels $\mylattice$ will be simply $\{l,h\}$, with $l \leq h$, where $l$ stands for ``low'' or ``public'', and $h$ stands for ``high'' or ``secret'' as also done in \cite{Focardi'02,Focardi-el'05,Boudol-Castellani'02}. 
The current work could be extended to a more general notion of lattice, however we argue that from a semantics and security point of view a more general notion of lattice would not give more expressiveness. 
In fact, in the semantics definition of non-interference we express the fact that public action cannot have any form of casual dependency from secret actions. 
This means that in a general lattice, actions below a certain security level are considered of public domain, and all the action above a given security level must be protected. 
That means in actual fact that it is sufficient to consider a collapsed lattice with two security levels only.

Before proceeding to the definition of the security relation we fix some notation.
\begin{definition}[Notation]
\begin{itemize}
\item We write $\procgenhigh$ for the subset of process that have prefixes with type $h$ only. 
\item We write $(\New A)\,P$ where $A$ is a {\em set of names} for the restriction in $P$ of all the names present in $A$.
\item
We write $(\New \highnames)\,P$ to to mean that we restrict {\em all} the names that have security level $h$ in the process $P$.
\end{itemize}
\end{definition}

The first definition of equivalence-based non interference uses the definition of weak bisimilarity directly.
\begin{definition}[\bndc] Let $P$ be a process.
 $P$ is said to be {\em secure}, $P\in \bndc$, if for every process $\Pi \in
\procgenhigh\,$, $\,(\New \highnames)(P \mid \Pi) 
\bisimilarity (\New \highnames) P$.
\end{definition}
The {\bndc} requires that high level actions present in the process $\Pi$ have have no effect on the execution of $P$.

Clearly any process $P$ which does not contain high names is secure. 
In fact, we have on one side $(\New \highnames)(P \mid \highprocesses) 
\approx P \mid (\New \highnames)( \highprocesses) $
where $(\New \highnames)( \highprocesses) \approx \nil$,
and on the other side $(\New \highnames)P \approx P $. Any process $P$ which contains only high names is secure, since all processes can only perform $`t$ actions.
Insecurity may appear when a high name is sequentially followed by a low name in $P$, because in this case the execution of $(\New \highnames) P$ may block on the high name (if this is reachable), making the low name unreachable, while it is always
possible to find a high process $\Pi$ that makes the low name
reachable in $(\New \highnames)(P \mid \Pi)$. 
Typical examples of insecure processes of this kind are $\In{a_h}(x).\Out{ b_l}< e>$ and 
$\In{a_h}(e).\Out{b_l}<e>$.
These examples show that the {\bndc} does not distinguish whether a low level action comes after an input or an output.
Quite surprisingly, insecurity appears when a high name is in conflict with a low name in $P$, that is, when they occur in different branches of a choice, as in the process $ \In a_h(x) + \Out b_l<e>$. 
It is disputable if this process should be considered insecure since the low and high level actions are independent. 
Finally, the process: $ \In{a_h}(x).\Out{b_l}<e> + \Out{b_l}<e> $ is secure.
 
As argued in~\cite{Focardi'02,Focardi-el'05}, Bisimulation-based Non-Deducibility on Compositions is not strong enough to deal with dynamic contexts. 
A strengthening of this notion, called Persistent Bisimulation-based Non-Deducibility on Compositions (\pbndc) was therefore proposed in~\cite{Focardi'02}.
We shall adopt this notion as the starting point for our study. 

To define \pbndc, a new kind of transition \raise2pt\hbox{$\wlts{\uptohigh{`a}}$} is introduced, defined as follows for any $`a\in \Act$.

\begin{definition}
The relation $\wlts{\uptohigh{a}}$ is defined as $\wlts{\hat{`a}} \union \ltss{`t}$ when $\subj{`a} \in\highnames $, or in the usual manner when $a$ is a low level action.
\end{definition}
The definition of weak bi-simulation up-to-high used the new relation in the definition.
 \begin{definition}[Weak bi-simulation up-to-high]
A symmetric binary relation $\Bisi \subseteq \Proc\times \Proc $ is a {\em weak bisimulation up-to-high} if an only if $P \Bisi Q$ implies that, for all $a \in \Act$:
\begin{itemize}
 \item whenever $P \lts{`a}P'$ then there exists $Q'$ such that $Q \wlts{\uptohigh{`a}} Q'$ and $P' \Bisi Q'$.
\end{itemize}
Two processes $P,Q$ are {\em weakly bisimilar up-to-high}, written $P \bisimilarityuptohigh Q$, if $P \Bisi Q$ for some weak bisimulation up-to-high $\Bisi$.
 \end{definition}
In other words, when a process makes a high-level action, could be matched by any number of $`t$-action. This definition abstracts away from high level actions.

 \begin{definition}[\pbndc]
$P $ is said {persistently secure}, $P \in \pbndc$ if $ (\New \highnames) P \bisimilarity_{\highnames} P $.
 \end{definition}

It has been shown in~\cite{Focardi'02,Focardi-el'05} that {\pbndc} is strictly stronger than {\bndc} i.e.$\pbndc \subset \bndc$.
In fact, if $P$ is in the {\pbndc} amounts to requiring {\bndc} for all reachable states of $P$; this explains why it is called ``persistent''.
The example considered above for {\bndc} are also persistently secure; however the process: 
\[\In a_h(v).\In a_h(v).\Out b_l<r> + \Out b_l<r>
\] 
is secure but not persistently secure.

\section{A simple type system}\label{simple-type-system}
In this section we will adapt the type system as developed by Pottier~\cite{Pottier'02} for the {\PI}, to CCS. 
That type system was devised with the idea of defining the simplest possible types that would guarantee non-interference. 
In that paper, Pottier works mostly with the $`p$-calculus with replication and general choice. 
In particular, we simplify the original type system and we adapt the rule of replication to recursion and eliminate the rule $(\ptnorm)$ used to guarantee that all the prefixes in the choice have the same security level. 
Because the version of CCS used here has only guarded choice, the rule $(\ptnorm)$ is not longer necessary. 


We will now introduce the type system: it assigns {\em security levels} to channels in processes. 

Security levels are elements $`s,`t$ of a lattice $\lattice{L}{\leq}$: a flow from level $`s$ to level $`t$ is authorised if and only if $`s\leq`t$.
We use $\meet$ and $\join$ for the operations of, respectively, meet and join on this lattice.

Type judgements for processes have then the form $\DerFT `G |- P : {\ell} $ which informally means that the process $P$ can be inferred from the environment $`G$ at security level $\ell$, where $\ell $ is a meta-variable ranging over the security lattice.


\begin{definition}[Type Assignment]
A {\em type environment} $`G$ is a mapping from channel names to security levels such that $`G(a) = `G(\overline{a})$; we write $a{:}\ell \in `G$ whenever $`G(a) = \ell$. We naturally extend the mapping to prefixes $`a$ by $`G(`a) =\subj{`a}$. 

The assignment of (security) types to processes is defined via the following natural deduction system.
\[\begin{array}{rl}
(\ptnil): & 
	\prove{\DerFT `G |- {\Nil} : {\ell} }

\\[3mm]
(\ptsub): &
	 \prove[\ell'\leq \ell]
	 	{ \DerFT `G |- P : {\ell} }
		{ \DerFT `G |- P : \ell' } 
\\[3mm]
(\ptcom): &
	\prove{ \DerFT `G |- P : {\ell} \quad \DerFT `G |- Q : {\ell} }
		{ \DerFT `G |- P\Par Q : {\ell} }
		\end{array}
\quad\quad
\begin{array}{rl}
(\ptrec): &
	\prove[\Agent{A}{x} =P]
		{ \DerFT `G, x_1{:}\ell_1, \ldots, x_n{:}\ell_n |- P : {\ell} } 
		{ \DerFT `G, b_1{:}\ell_1, \ldots, b_n{:}\ell_n |- \Agent{A}{b} : {\ell} }
\\[3mm]
(\ptsum): &
\prove	{ \DerFT `G |- `a_i.P_i : {\ell} \quad `G(`a_i)=\ell \quad (\forall i\in I)}
	{ \DerFT `G |- \Choice{i}{I}`a_i.P_i : {\ell} }
\\[3mm]
(\ptres): &
	\prove{ \DerFT `G,a{:}\ell' |- P : {\ell} }
		{ \DerFT `G |- (\New a_{\ell'})\,P : {\ell} }
\end{array}\]
\[\begin{array}{rlcrl}
\end{array}\]

\end{definition}

\begin{definition}
$P$ is \emph{typeable} if $\DerFT `G |- P : {\ell} $ for some $`G$ and $\ell$.
\end{definition}

Clearly not all processes are typeable. For instance $\In{a_h}(x).\Out{b_h}<e>$ is not typeable. 
Here the difference between the type system and the general typed language as defined in this paper is made clear. 
The type language does not impose any constraint on the construction of processes. 
Thus, the process $\In{a_h}(x).\Out{b_h}<e>$ is a legal term according to our syntax, but it is not possible to find an environment $`G$ such that will assign to the process $\In{a_h}(x).\Out{b_h}<e>$ a type $\ell$.

The following theorem states that no matter how the process behaves, there will be no leakage of sensitive data, since types are preserved by reductions.

\begin{Theorem}[Subject reduction] 
If $\DerFT `G |- P : {\ell} $ and $P \lts{`a} P'$ then $\DerFT `G |- P' : {\ell} $.
\end{Theorem}

\begin{proof}
By induction on the inference of $\DerFT `G |- P : {\ell} $.
\end{proof}

In this section we analyse the relationship between the 'simple type system' developed by Pottier \cite{Pottier'02} and {\pbndc} \cite{Focardi-el'05}. 
We shall see that every typeable process according to Pottier's type system is secure according to the \pbndc. 



\begin{proof} By induction on the inference of $\DerFT `G |- P : {\ell} $.
\end{proof}
\begin{Theorem}\label{soundness} 
If $P$ is typeable, then $P \in \pbndc$.
\end{Theorem}

The reverse of the above theorem is not true. 
In fact, $a_h.b_l + b_l \in {\pbndc} $ while this process cannot be typed in the type system above. 
We conclude that if $P$ is typeable then it is persistently secure and secure. By the examples presented in this paper, not all secure processes are typeable or persistently secure.

 Also $\approx_\highnames$ is not preserved by parallel composition
on arbitrary programs, as shown by the following example where $P_i \bisimilarityuptohigh Q_i$
for $i =1,2\,$ but $P_1\mid P_2 \not \bisimilarityuptohigh Q_1\mid Q_2$.
Take
\[ P_1 = \In{a_h}(x) \quad Q_1 = \nil \quad P_2 = Q_2 = 
(\New \Out{b_h}<e> \mid \In{b_h}(x) ) ( \Out{c_l}<e'>. + \Out{a_h}<e''>).\] 
Clearly $\Out{c_l}<e'> + \Out{a_h}<e''>$ is not typeable since in the sum only prefixes at the same security level are allowed. 
This means that for untyped processes the {\pbndc} is not closed under arbitrary contexts, which makes compositional reasoning quite difficult. It is an open question --which we leave for future work-- whether {\pbndc} is closed under typed contexts.

In this section we have shown that the 'simple type system ' has a natural correspondence in the {\pbndc}. A type system gives an automatic way to guarantee the bsence of leakage in programs. This is the main advantage of the type system over semantics based notions of non-interference.

\section{Asymmetric type system for CCS }\label{refined-type-system}
The `simple type system' imposes as security discipline such after high level action only low-level actions can follow. 
In other words, the type systems guarantees that there is not causal dependency from high level action to low-level actions.
We argue that there is a difference between the action performed by an input and an output.
Consider the example of two systems, where the first one simply emits signals of acknowledgements to both high and low.
\[ P(ack) = \Out{ack_h}<e>.\Out{ack_l}<e'>.P\]
The second system is a system that first reads from a secret database and then outputs the outcome.
\[ Q(ack) = \In{read}(x). \Out{wait_l}<.e>.\Out{write_l}<x>.P\]
Clearly for $P$ is makes no difference in which order the high-level and the low-level actions take place. In no way $ack_l$ can reveal anything about $ack_h$ since the value of the outputs are independent. However, the situation is radically different for $Q$.
After an high-level input, information can be {\em leaked} to an insecure level via a low-level output as defined in $Q$. Therefore, it is vital that after a high input, a low level output action is not permitted. The type system we present in the next section is a refinement of the simple type system, and distinguishes between input and output. It allows low-level actions after a high-level output under the assumption that high level outputs are not sensitive actions. On the other end, it not possible to perform a low-level action after an input as in the simple type system.

The types developed in this section are inspired by those of~\cite{Boudol-Castellani'02}: they record both the {\em reading level} of processes (as the maximal level of their input channels) and their {\em writing level} (the minimal level of their output channels). 




Type judgements for processes have the form $\Der `G |- P : (B,`s, `t) $, where $`s$ is an {\em upper bound} for the level of input channels of $P$, and $`t$ is a {\em lower bound} for the level of its output channels. 


 
Notice that we have a case of \emph{leakage} whenever an output takes place of a level \emph{lower} than the level of one of the inputs.
Therefore, a flow from level $`s$ to level $`t$ is authorised if and only if $`s\leq`t$.
In line with this intuition, subtyping for processes is covariant in its second argument and contra-variant in its third argument.

\begin{definition}A \emph{ type environment} $`G$ is a mapping from channel names to security levels such that $`G(a) = `G(\overline{a})$:
we write $a{:}\ell \in `G$ whenever $`G(a) = \ell$. 
 
Security type assignment on processes is defined by the following natural deduction system; 

\[\begin{array}{rl}
(\ruletypenil): &\quad
\prove	{ \Der `G |- {\Nil} : (\notout,\typebot,\typetop) }
\Comment{
\[\begin{array}{rl}
\hbox to 2cm{~} & \hbox to 10cm{~} \\[-24pt]
(\ruletypeinputone): &
\prove	[`r \leq `t]
	{ \Der `G,a{:}`r |- P : (\notout,`s,`t) }
	{ \Der `G |- \In a_{`r}(v).P : (\notout,`r \join `s,`t) }
\end{array}\]
\[\begin{array}{rl}
\hbox to 2cm{~} & \hbox to 10cm{~} \\[-16pt]
(\ruletypeinputtwo): &\quad
\prove	[`r \leq `t]
	{ \Der `G,a{:}`r |- P : (\yesout,`s,`t) }
	{ \Der `G |- \In a_{`r}(v).P : (\yesout,`r \join`s ,`t) }
\end{array}\]
\[\begin{array}{rl}
\hbox to 2cm{~} & \hbox to 10cm{~} \\[-16pt]
(\ruletypeoutputone): &
\prove	[`s \leq `r]
	{ \Der `G,a{:}`r |- P : (\notout,`s,`t) }
	{ \Der `G |- \Out a_{`r}<v>.P : (\yesout,`s,`r \meet`t) } 
\end{array}\]
\[\begin{array}{rl}
\hbox to 2cm{~} & \hbox to 10cm{~} \\[-16pt]
(\ruletypeoutputtwo): &
\prove	[`s \leq `r \meet`t]
	{ \Der `G,a{:}`r |- P : (\yesout,`s,`t) }
	{ \Der `G |- \Out a_{`r}<v>.P : (\yesout,`s,`r \meet`t) } 
\end{array}\]
}
\\[3mm] 
(\ruletypeinput): &\quad
\prove	[a{:}`r \in `G, `r \leq `t]
	{ \Der `G |- P : (B,`s,`t) }
	{ \Der `G |- \In a_{`r}(v).P : (B,`r \join`s ,`t) }
\\[3mm] 
(\ruletypeoutput): &
\prove	[a{:}`r \in `G, `s \leq `r]
	{ \Der `G |- P : (B,`s,`t) }
	{ \Der `G |- \Out a_{`r}<e>.P : (\yesout,`s,`r \meet`t) } 
\\[3mm] 
(\ruletypepar): &
\prove	[`s_1 \leq `t_2~\&~`s_2 \leq `t_1]
	{ \Der `G |- P : (\gentype_1,`s_1,`t_1) 
	 \quad 
	 \Der `G |- Q : (\gentype_2,`s_2,`t_2) 
	}
	{ \Der `G |- P\Par Q : (\gentype_1 \wedge \gentype_2, `s_1 \join `s_2,`t_1 \meet`t_2) } 
\\[3mm] 
(\ruletypesum): &
\prove	{ \Der `G |- a_i. P_i : (\gentype_1,`s,`t)
	 \quad
	}
	{ \Der `G |- \sum_{i\in I} `a_i.P_i : (\gentype_1 \wedge \gentype_2, `s,`t) } 
\\[3mm] %
(\ruletyperestr): &
\prove	{ \Der `G,a{:}`r |- P : (\gentype,`s,`t) }
	{ \Der `G |- (\New{a_{`r}})\,P : (\gentype,`s,`t) } 
\end{array}\]
\[\begin{array}{rl}
(\ruletyperec): &
\prove	[\Agent{A}{x} \Def P]
	{ \Der `G,x_1{:}`s_1, \ldots,x_n{:}`s_n |- P : (\gentype,`s,`t) }
	{ \Der `G,a_1{:}`s_1, \ldots,a_n{:}`s_n |- \Agent{A}{a} : (\gentype,`s,`t) } 
\\[3mm] 
(\ruletypesub): &
\prove	[`s_1 \leq `s_2 \leq `t_2 \leq `t_1]
	{ \Der `G |- P : (\gentype,`s_1,`t_1) }
	{ \Der `G |- P : (\gentype,`s_2,`t_2) }
\end{array}\]

\end{definition}

The side-conditions on levels guarantee than the input level never becomes bigger that the output level.
For instance, 
a program of type $(\bot, \top)$ is guaranteed to not perform any input on a high channel nor any output on a low channel.

Our type system aims to capture the property that in the presence of an output, which is the means for an observer to deduce implicit flows in the program, any previous input has to be done at a lower level.
Thus, a secure programs is one that for instance never emits an output. A secure program is also one that after every input emits only output of higher level, as expressed by the type $(`s,`t)$ where $`s \leq `t$.
These property are preserved by subject reduction as shown.

\begin{Proposition}[Subject Reduction] 
If $\Der `G |- P : (\gentype`,`s,`t)$ and $P \lts{`a}P'$ then $\Der `G |- P' : (\gentype,`s,`t)$.
.\end{Proposition}
\begin{proof}By induction on $\Der `G |- P : (\gentype,`s,`t)$.
\end{proof} 
We report in this section some examples of processes to show the power of discrimination of our type system. Some examples are taken from \cite{Focardi-el'05}.

\begin{Example}

Consider $\In a_h(r).\Out b_l<r>$, $\Out a_h<v>.\Out b_l<r>$, and $\In a_h(v).\In b_l(r)$.
None of these processes is considered secure under \bndc.
 \[ 
(\New \highnames) ( \Out a_h<v>.\Out b_l<r> \mid \highprocesses) \notbisi (\New \highnames) ( \Out a_h<r>.\Out b_l<r> ) 
 \]
This process is not secure because a high level action, either input or output, precedes a low level action.
Our type system distinguishes between either high-level input or high-level output performed before a low-level action.
We first consider $\Out a_h<v>.\Out b_l<r>.\Nil$.
\[
\prove	
	{ \prove 
		{ \Der `G , a{:}h,b{:}l |- {\Nil} : (\notout,l,h) }
		{ \Der `G, a{:}h,b{:}l |- \Out b_l<r>.{\Nil} : (\yesout,l,l) }
	}
	{ \Der `G, a{:}h,b{:}l |- \Out a_h<v>.\Out b_l<r>.{\Nil} : (\yesout,l,l) }
\]

We now consider $\In a_h(v).\In b_l(r).\Nil$.

\[
\prove	[h\not \leq l]
	{ \prove 
		{ \Der`G, a{:}h,b{:}l |- {\Nil} : (\notout,l,h) } 
		{ \Der `G, a{:}h,b{:}l |- \Out b_l<r>.{\Nil} : (\yesout,l,l) }
	}
	{ \Der `G, a{:}h,b{:}l |- \In a_h(v).\Out b_l<r>.{\Nil} : (\yesout,?,?) }
\]
In our type system, this process is not secure.

\end{Example}

\begin{Example}
We consider now the process $\Out a_h<v>.\Out a_h<v>.\Out b_l<r> + \Out b_l<r>$ is secure in {\bndc} but not in \pbndc:
\[
\prove	
	{ \prove 
		{ \Der `G, a{:}h,b{:}l |- \Out a_h<v>.\Out b_l<r> : (\yesout,l,l) }
		{ \Der `G, a{:}h,b{:}l |- \Out a_h<v>.\Out a_h<v>.\Out b_l<r> : (\yesout,l,l) }
	 \quad
	 \Der `G , a{:}h,b{:}l |- \Out b<r>.{\Nil} : (\yesout,l,l) 
	}
	{ \Der `G, a{:}h, b{:}l |- \Out a_h<v>.\Out a_h<v>.\Out b_l<r> + \Out b_l<r> : (\yesout,l,l) }
\]

The processes
$\In a_h(v).\In a_h(v).\Out b_l<r> + \Out b_l<r>$ would be still secure in {\bndc} but not in \pbndc, while clearly this process is not typeable in our type system.
\end{Example}

\begin{Example} 
The process $\In{a_l}(x) + \Out{b_h}<e>$ which is not included in neither the {\bndc} nor in the {\pbndc} is secure according to the current type system.
\[ \prove 
	{ \Der `G,a{:}l,b{:}h |- \In a_l(x) : (\yesout,l,l) 
	 \quad
	 \Der `G,a{:}l,b{:}h |- \Out b_h<r> : (\yesout,l,l) 
	}
	{ \Der `G,a{:}l,b{:}h |- \In a_l(x) + \Out b_h<e> : (\yesout,l,l) }
\]
\end{Example}
Clearly by the examples presented above, if $\Der `G |- P : (\gentype,`s,`t)$ then $P \not \in \bndc$ nor $P \not \in\pbndc$. 
It remains an interesting question what equivalence relation could be characterised by this type system.

We propose here a candidate which is a variation on both the {\bndc} and {\pbndc} and we leave for future work to analyse the formal relationship with the type system.

 \begin{definition}[Refined Weak Bisimulation up-to-high]
A symmetric binary relation $\Bisi \subseteq \Proc\times \Proc $ is a {\em refined weak bisimulation up-to-high} if and only if $P \Bisi Q$ implies that, for all $a \in \Act$ either 
\begin{itemize}
\item if $P \lts{`a}P'$ and $`a = \overline{a}e$ then there exists $Q'$ such that $Q \wlts{\uptohigh{`a}} Q'$ and $P' \Bisi Q'$; or 
\item if $P \lts{`t}P'$ then there exists $Q'$ and a channel name $a_\rho$ and a value $e$ such that $`a = \overline{a}e $ and $Q \wlts{\uptohigh{`a}} Q'$ and $\subj{`a} = a_\rho, \rho= h$ and $P' \Bisi Q'$ or 
$Q \wlts{\uptohigh{`t}} Q'$ and and $P' \Bisi Q'$; or 
\item f $P \lts{`a}P'$ and $`a = ae$ then there exists $Q'$ such that $Q \wlts{`a} Q'$ and $P' \Bisi Q'$.
\end{itemize} 
Two processes $P,Q$ are {\em refined weakly bisimilar up-to-high}, written $P \approx^I_\highnames Q$, if $P \Bisi Q$ for some refined weak up-to-high bisimulation $\Bisi$.
 \end{definition}

The definition of Refined Weak bi-simulation up-to-high aims to distinguish between inputs and outputs. It is designed with the principles described below.
\begin{itemize}
\item High-level outputs can be matched by weak transition of the same name or any sequence of $`t$-actions.
\item $`t$-actions can be matched either by any sequence of $`t$-actions.
or by weak transitions of high-level output.
\item Input can be matched only by weak transitions of the same name regardless the security level.
\end{itemize}
\begin{definition}
Let $A \subset \highnames$. We define $\Phi_A = \prod_{a \in A} \In{a_h}(z) \mid \Phi_A$.
\end{definition}
 \begin{definition}[\wbndc]
$P $ is said to be \emph{generally secure}, $P \in \wbndc$ if 
$ (\New \highnames) (P \mid \Phi_{\fn{P}}) \bisimilarity^I_{\highnames} P $.
 \end{definition}
 The process $\Phi_A$ generate high-level input of the channels contained in $A$.

According to this definition of the process $\In a_h(v).\In a_h(v).\Out b_l<r> + \Out b_l<r>$ would not be considered generally secure. 
In fact, $(\New \highnames) (\In a_h(v).\In a_h(v).\Out b_l<r> + \Out b_l<r> \mid \Phi_{a})$ the left-hand of the sum is blocked. Let's consider:
\[ E= (\New \highnames) (\In a_h(v).\In a_h(v).\Out b_l<r> + \Out b_l<r> \mid \Pi) \quad \quad G =(\In a_h(v).\In a_h(v).\Out b_l<r> + \Out b_l<r> \]

Assume that $G \lts{ae} G'$ then $E$ can only stay put. It is not difficult to show that $ G \bisimilarity^I_{\highnames} E $ does not hold. If $E \lts{br} \Nil$ then $G$ cannot match it with any low-level action. 

On the other hand $\Out a_h<v>.\Out a_h<v>.\Out b_l<r> + \Out b_l<r>$ is generally secure.


\section*{Conclusions}\label{conclusions}
In this paper we have considered two different approaches to non-interference, namely a static approach via a simple type system and a semantic approach via \pbndc. 
We have shown that the `simple type system' is correct with respect to \pbndc, yet not complete.
 We have also defined a new type system that distinguishes between information flows from inputs and outputs. Information flow from high -level outputs to low-level channels is considered safe in the new type system.
 We defined also the Refined Weak Bi-simulation up-to-high which aims to characterise the refined type system.
 
As far as future work is concerned it would be interesting to relate typed language-based notion of non-interference with a process algebraic approach similarly to the work done in~\cite{Focardi-el'05} for typed languages. 
In particular, it would be interesting to consider the type system of Volpano \cite{Volpano-el'96} or Boudol and Castellani \cite{Boudol-Castellani'02} to define a type system in the process language that preserves that notion of non-interference.

\subsection*{Acknowledgements}
We gratefully acknowledge Ilaria Castellani for many useful discussions and for having pointed out mistakes in the proof of Theorem \ref{soundness} and for suggesting corrections. 
She also observed that the {\pbndc} is not closed under general contexts and provided the example reported in this paper.
We gratefully acknowledge the Group MIMOSA at INRIA Sophia-Antipolis where this work was initially conceived for their hospitality during 2005-2006.

\bibliographystyle{plain}

\end{document}